\documentclass[fleqn,usenatbib]{mnras}

\usepackage{newtxtext,newtxmath}

\usepackage[T1]{fontenc}
\usepackage{ae,aecompl}

\usepackage{graphicx}	
\usepackage{amsmath}	
\usepackage{amssymb}	
\usepackage{xspace}        

\title[Nuclear radio activity in PDS~456]{The nearby extreme accretion and feedback system PDS~456: finding a complex radio-emitting nucleus} 

\author[J. Yang et al.]
{Jun Yang,$^{1}$\thanks{E-mail: jun.yang@chalmers.se}
Zsolt Paragi,$^{2}$
Emanuele Nardini,$^{3,4}$
Willem A. Baan,$^{5}$
Lulu Fan,$^{6,7,8}$
\newauthor
Prashanth Mohan,$^{9}$
Eskil Varenius$^{1,10}$
and Tao An$^{9}$
\\
$^{1}$Department of Space, Earth and Environment, Chalmers University of Technology, Onsala Space Observatory, SE-439 92 Onsala, Sweden \\
$^{2}$Joint Institute for VLBI ERIC (JIVE), Postbus 2, NL-7990 AA Dwingeloo, the Netherlands \\
$^{3}$Dipartimento di Fisica e Astronomia, Universit\`a di Firenze, via G. Sansone 1, I-50019 Sesto Fiorentino, Firenze, Italy \\
$^{4}$INAF – Osservatorio Astrofisico di Arcetri, Largo Enrico Fermi 5, 50125 Firenze, Italy \\
$^{5}$Netherlands Institute for Radio Astronomy ASTRON, NL-7991 PD Dwingeloo, the Netherlands \\
$^{6}$CAS Key Laboratory for Research in Galaxies and Cosmology, Department of Astronomy, University of Science and Technology of China, 230026 Hefei, China \\
$^{7}$School of Astronomy and Space Sciences, University of Science and Technology of China, 230026 Hefei Anhui, China \\
$^{8}$Shandong Provincial Key Lab of Optical Astronomy and Solar-Terrestrial Environment, Institute of Space Science, Shandong University, 264209 Weihai, China \\
$^{9}$Shanghai Astronomical Observatory, Key Laboratory of Radio Astronomy, Chinese Academy of Sciences, 200030 Shanghai, China \\
$^{10}$Jodrell Bank Centre for Astrophysics, The University of Manchester, Oxford Rd, Manchester M13 9PL, UK
\\
}
\date{Accepted 2020 XXX. Received 2020 YYY; in original form 2020 ZZZ}

\pubyear{2018}

\begin{document}
\label{firstpage}
\pagerange{\pageref{firstpage}--\pageref{lastpage}}
\maketitle

\begin{abstract}
When a black hole accretes close to the Eddington limit, the astrophysical jet is often accompanied by radiatively driven, wide-aperture and mildly relativistic winds. Powerful winds can produce significant non-thermal radio emission via shocks. Among the nearby critical accretion quasars, PDS~456 has a very massive black hole (about one billion solar masses), shows a significant star-forming activity (about seventy solar masses per year) and hosts exceptionally energetic X-ray winds (power up to twenty per cent of the Eddington luminosity). To probe the radio activity in this extreme accretion and feedback system, we performed very-long-baseline interferometric (VLBI) observations of PDS~456 at 1.66~GHz with the European VLBI Network (EVN) and the enhanced Multi-Element Remotely Linked Interferometry Network (e-MERLIN). We find a rarely-seen complex radio-emitting nucleus consisting of a collimated jet and an extended non-thermal radio emission region. The diffuse emission region has a size of about 360~pc and a radio luminosity about three times higher than the nearby extreme starburst galaxy Arp~220. The powerful nuclear radio activity could result from either a relic jet with a peculiar geometry (nearly along the line of sight) or more likely from diffuse shocks formed naturally by the existing high-speed winds impacting on high-density star-forming regions.  
\end{abstract}




\section{Introduction}
\label{sec1}
Accreting supermassive black holes (SMBHs) can provide mechanical feedback on their host galaxies via launching jets and winds. Jets are collimated relativistic outflows emitting synchrotron emission \citep[e.g.][]{Blandford2019}. Winds are non-jetted, relatively wider angled outflows with a relatively low speed of $\la$0.3~$c$ \citep[e.g.][]{Tombesi2016} and are mainly traced by optical and X-ray spectroscopic as well as radio continuum observations 
\citep[e.g.][]{Panessa2019}. When the SMBHs have accretion rates approaching the Eddington limit, they may produce not only jets \citep[e.g.][]{Giroletti2017, Yang2019, Yang2020a} but also extremely powerful winds \citep[e.g.][]{Nardini2015, Tombesi2015, Fiore2017}. 
These winds can sweep out the interstellar medium \citep[e.g.][]{Zakamska2014}, produce radio-emitting shocks \citep[e.g.][]{Nims2015}, contribute the extragalactic gamma-ray background \citep[e.g.][]{Lamastra2017} and quench star formation \citep[e.g.][]{Kormendy2013, Fiore2017}. Observing this complex nuclear radio activity with the very-long-baseline interferometric (VLBI) imaging technique can provide information to constrain physical properties in the extreme accreting environment and help understand feedback with the host galaxy. 

The quasar PDS~456 (IRAS~17254$-$1413) at a redshift $z = 0.184$ is the most luminous quasar in the local ($z < 0.3$) Universe \citep{Torres1997, Simpson1999}. The quasar has a bolometric luminosity $L_\mathrm{bol} \sim 10^{47}$~erg\,s$^{-1}$ \citep{Reeves2000} comparable to its Eddington luminosity $L_{\rm Edd} \sim 1.3~\times~10^{47}$~erg\,s$^{-1}$ \citep{Nardini2015} and a disk geometry close to face-on \citep{Yun2004, Bischetti2019}. It is also an ultraluminous infrared (IR) galaxy (ULIRG) with a luminosity of $L_{\rm FIR} = 1.3 \times 10^{12}$~L$_{\sun}$ \citep{Yun2004}. Compared to the known nearby critical accretion systems, its central SMBH has the highest mass, $M_{\rm bh}~=~10^{9.2\pm0.2}$~M$_{\sun}$ \citep[supplementary materials,][]{Nardini2015}.

As a critical accretion system, PDS~456 is of great interest for probing powerful multi-phase winds and outflows. It hosts a quasi-spherical mildly relativistic ($\la$0.3~$c$) X-ray wind with a very high kinetic energy $\sim$0.2$L_{\rm Edd}$ \citep{Nardini2015}. The power of X-ray winds positively correlates with the X-ray luminosity, and thus these winds are likely radiation pressure driven \citep{Matzeu2017}. From soft X-ray \citep{Boissay-Malaquin2019, Reeves2020} to ultra-violet \citep{OBrien2005, Hamann2018} bands, there are also more reports of highly blue-shifted absorption lines at velocities consistent with the hard X-ray winds. Moreover, the CO (3--2) emission line observations at $\sim$1~mm show that there exists not only some spatially extended molecular outflows up to $\sim$5~kpc but also high-velocity ($\sim$800~km\,s$^{-1}$) compact outflows in the nucleus \citep{Bischetti2019}. 

Based on the multi-wavelength spectral energy distribution (SED), PDS~456 is a radio-quiet analogue of the well-known radio-loud blazar 3C~273 \citep{Yun2004}. The radio counterpart of PDS~456 is a steep-spectrum source \citep{Yun2004} with a flux density of 25~$\pm$~4~mJy at 1~GHz and a spectral index of $\alpha = -$0.84~$\pm$~0.11 \citep{Yang2019}. Besides the intense active galactic nucleus (AGN), its SED in the far IR shows evidence for significant star-forming activity in its host galaxy \citep{Yun2004}. The radio emission is likely dominated by the AGN activity \citep{Yun2004}. The pilot high-resolution VLBI observations at 5~GHz found a faint jet structure consisting of a few components on the deca-pc scale in the nuclear region, while failed to image some very diffuse radio structure \citep[$\sim$70 per cent of its total flux density,][]{Yang2019}. To get a complete view on the potential large scale radio activity which may owe to an episodic jet, and understand nuclear starbursts and/or wind shocks, we performed new VLBI imaging observations at a suitable frequency of 1.66~GHz with the European VLBI Network (EVN) plus the enhanced Multi-Element Remotely Linked Interferometry Network (e-MERLIN). 

This paper is organised as follows. We describe the VLBI observations and the data reduction in Section~\ref{sec2} and present the imaging results of PDS~456 in Section~\ref{sec3}. We discuss star-formation rate based on the IR SED, physical characteristics of the observed complex nuclear radio activity and potential identification with the jet, starburst and shock activity, and the accretion rate in Section~\ref{sec4}. Throughout the paper, a standard $\Lambda$CDM cosmological model with $H_{\rm 0}$~=~71~km\,s$^{-1}$\,Mpc$^{-1}$, $\Omega_{\rm m}$~=~0.27, $\Omega_{\Lambda}$~=~0.73 is adopted; the images then have a scale of 3.1~pc\,mas$^{-1}$.

\section{Observations and data reduction}
\label{sec2}

\begin{figure}
\centering
\includegraphics[width=\columnwidth]{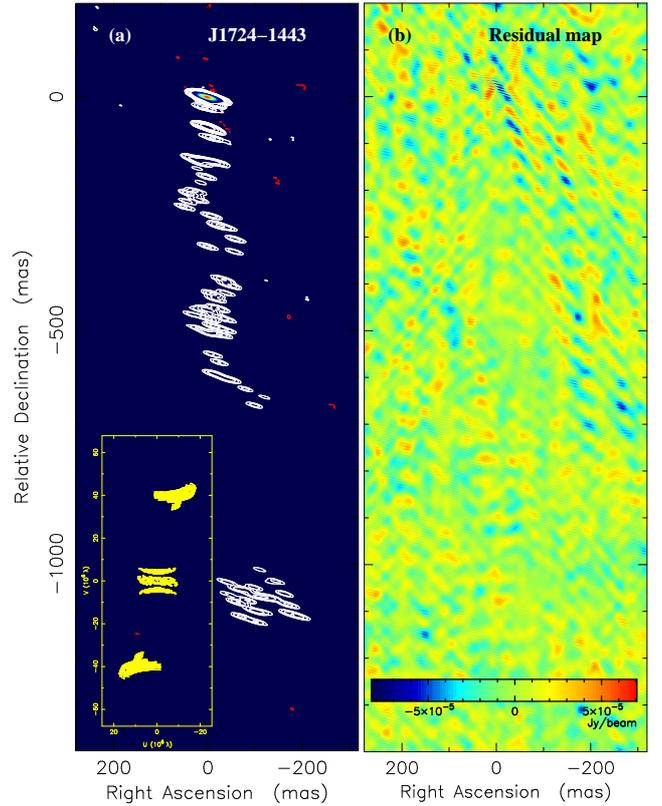}  \\
\caption{
The point-source model fitting results of the visibility data of the calibrator J1724$-$1443. (a) The high-dynamical range map of its complex jet structure. The full width at half maximum (FWHM) of the synthesised beam is 29.6~mas~$\times$~6.6~mas at 73.9$\degr$. The contours are 0.05~$\times$~($-$1, 1, 2, ..., 64)~mJy\,beam$^{-1}$.  The image peak is 229~mJy\,beam$^{-1}$. With respect to the image noise level 0.017~mJy\,beam$^{-1}$, the dynamic range reaches 13\,740. The inset plots the non-optimal ($u$, $v$) coverage. (b) The noise distribution in the residual map. The display range is between $-$0.083 to $+$0.070 mJy\,beam$^{-1}$.}
\label{fig1}
\end{figure}

\begin{table}
\caption{Summary of the VLBI observations of PDS~456 at 1.66 GHz. }
\label{tab1}
\begin{tabular}{cc}
\hline\hline
Observing date and time (UT)          & Participating stations    \\     
\hline     
2018 Mar 28, 02h30m--08h30m   & JbWbEfMcOnTrHh               \\
2020 Jan 22, 06h30m--13h00m   & JbWbEfOnTrHhIrSrKnDaPiCmDe   \\
\hline
\end{tabular}
\end{table}

Table~\ref{tab1} gives a summary of our VLBI observations at 1.66~GHz. At first, we observed PDS~456 with the EVN on 2018 March 28. Owing to a severe instrumental problem with the digital filters, the target was re-observed with the EVN plus the e-MERLIN on 2020 January 22. In the two experiments, the participating stations were Jodrell Bank MK II (Jb), Westerbork single antenna (Wb), Effelsberg (Ef), Medicina (Mc), Onsala (On), Toru\'n (Tr), Hartebeesthoek (Hh), Irebene (Ir), Sardinia (Sr), Knockin (Kn), Darnhall (Da), Pickmere (Pi), Cambridge (Cm), Defford (De).      

Both observations were carried out in the e-VLBI mode. The data transfer speeds were 1024~Mbps (16 sub-bands in dual polarisation, 16~MHz per sub-band, 2-bit quantisation) at the EVN stations and 512~Mbps (2 sub-bands in dual polarisation, 64~MHz per sub-band, 2-bit quantisation) at the e-MERLIN stations (Kn, Da, Pi, Cm and De). The data correlation was done in real time by the EVN software correlator \citep[SFXC,][]{Keimpema2015} at JIVE (Joint Institute for VLBI, ERIC) using an integration time of 1~s and a frequency resolution of 1~MHz.    

The observations of the faint source PDS~456 applied the phase-referencing observing technique. The pc-scale compact source J1724$-$1443 \citep{Petrov2006}, about 59~arcmin apart from our target, was used as the phase-referencing calibrator. Its J2000 position is RA~$= 17^{\rm h}24^{\rm m}46\fs96654$ ($\sigma_{\rm ra} = 0.1$~mas), Dec.~$= -14\degr43\arcmin59\farcs7609$ ($\sigma_{\rm dec} = 0.2$~mas) in the source catalogue 2016a from the Goddard Space Flight Centre (GSFC) VLBI group. The cycle time of the periodic nodding observations was about four minutes ($\sim$0.5~min for J1724$-$1443, $\sim$2.5~min for PDS~456, $\sim$1.0~min for two gaps). A very bright flat-spectrum radio quasar NRAO~530 \citep[J1733$-$1304, e.g.][]{An2013} was also observed to determine the instrumental phases and bandpass shapes.  

The data were calibrated using the National Radio Astronomy Observatory (NRAO) software package Astronomical Image Processing System \citep[\textsc{aips}, ][]{Greisen2003}. First, we reviewed the data carefully and flagged out off-source data and some very low-sensitivity data. Second, we ran a-priori amplitude calibration with properly smoothed antenna monitoring data (system temperatures and gain curves) or nominal system equivalent flux densities when these monitoring data were not available. Third, we updated the position of the antenna Sr with the \textsc{aips} task \textsc{clcor} (options \textsc{ANAX} and \textsc{ANTP}) according to the first geodetic VLBI measurements (project code EL054, reported by Leonid Petrov on the web\footnote{\url{http://astrogeo.org/petrov/discussion/el054/}}): J2000 epoch, position: ($X$, $Y$, $Z$)~=~($+$4865183.542, $+$791922.255, $+$4035136.024)~m, velocity: ($V_{\rm x}$, $V_{\rm y}$, $V_{\rm z}$)~=~($-$12.15, $+$19.29, $+$10.86)~mm\,yr$^{-1}$ and axis offset 0.031~m. The corrections are ($\Delta X$, $\Delta Y$, $\Delta Z$)~=~($+$0.6116, $-$0.1553, $-$1.0446)~m for the antenna position and $+$0.031~m for the antenna axis offset. Fourth, the general bootstrap phase calibrations were performed. We corrected the ionospheric dispersive delays according to the maps of total electron content provided by Global Positioning System (GPS) satellite observations, removed phase errors due to the antenna parallactic angle variations, aligned the phases across the subbands via running fringe-fitting with a two-minute scan of the NRAO~530 data. After these calibrations, we combined all the subbands, ran the fringe-fitting and applied the solutions to PDS~456 by interpolation. Finally, the bandpass calibration was performed. All the above steps were scripted in the ParselTongue interface \citep{Kettenis2006}.

We first imaged the phase-referencing source J1724$-$1443. The imaging procedure was performed through a number of iterations of model fitting with a group of delta functions, i.e., point source models, and the self-calibration in \textsc{difmap} \citep{Shepherd1994}. This is similar to the non-negative least-squares algorithm \citep[NNLS,][]{Briggs1995} in the image plane and thus can also allow users to achieve a high-dynamic-range map \citep{Yang2020b}. We re-ran the fringe-fitting and the amplitude and phase self-calibration in \textsc{aips} with the input source model made in \textsc{difmap}. All these solutions were also transferred to the data of PDS~456 by the linear interpolation. 

The total intensity and residual maps from the direct model fitting are shown in Fig.~\ref{fig1}. The calibrator J1724$-$1443 has a single-sided core-jet structure with a total flux density of 0.27~$\pm$~0.02~Jy. Its radio core has a flux density of 0.23~$\pm$~0.01~Jy and is used as the reference point in the phase-referencing calibration. There were 79 point source models used in the Stokes $I$ map. Both the Stoke $I$ and zero-flux density $V$ maps have almost the same off-source noise level, $\sigma_{\rm rms}$ = 0.017~mJy\,beam$^{-1}$. The noise distribution in the residual map of Stokes $I$ is quite uniform in particular in the on-source region, although there is only one station (Hh) on the long baselines. Because the method tried to use the minimum number of point sources and the ($u$, $v$) coverage is poor, the faint jet structure does not looks very smooth. The e-MERLIN data were excluded in the final image because they gave some low-level noise peaks ($\sim$0.15~mJy~beam$^{-1}$, $\sim$0.07 per cent of the peak brightness) in the residual map. 

We imaged the faint target PDS~456 without any self-calibration in \textsc{difmap}. First, we made a low-resolution \textsc{clean} map with the e-MERLIN and Jb. The target PDS~456 has a peak brightness of 7.6$\pm$0.2~mJy\,beam$^{-1}$ using natural weighting (beam FWHM: 506~$\times$~111 mas$^{2}$ in position angle 14$\degr$), and indicates a slightly resolved structure with a total flux density of 12.3$\pm$0.3~mJy and a de-convolved size of $\sim$123~mas. After the deconvolution of the major component with a proper window, there is a hint of two-sided jet-like structure at position angle (PA) of about $-$25$\degr$ and $+$145$\degr$ in the residual map but they are not bright enough ($<$5$\sigma$) for a confirmation as real features. This is followed up by making a high-resolution map with the EVN plus e-MERLIN data. We used $\sigma^{-2}$ ($\sigma$, data error) as the visibility data weight and the purely natural grid weighting to clean the diffuse emission (peak brightness: 0.23~mJy\,beam$^{-1}$) in the nuclear region. Owing to residual phase errors of the phase-referencing calibration, our final image has a noise level about a factor of two higher than the statistical value estimated in the zero-flux Stokes $V$ map.  

\section{EVN plus e-MERLIN imaging results}
\label{sec3}

\begin{figure}
\includegraphics[width=\columnwidth]{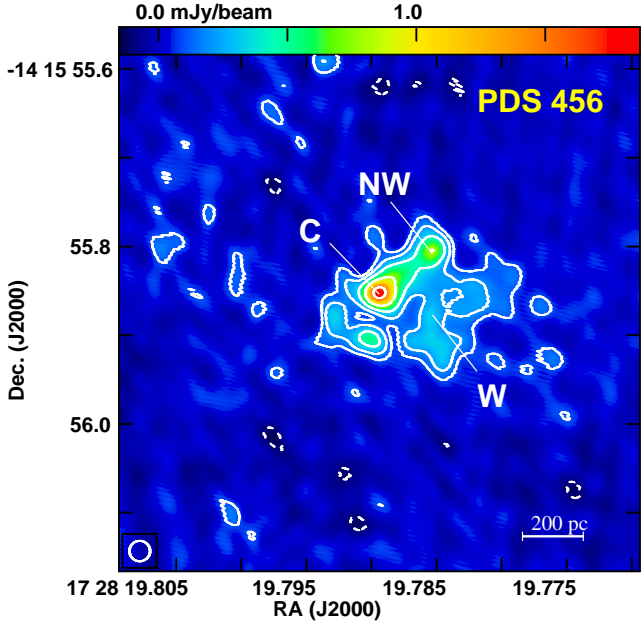}  \\
\caption{
The total intensity image of the quasar PDS~456 observed by the EVN plus e-MERLIN at 1.66~GHz. A circular restoring beam with FWHM~=~23~mas is used. The contours start from 0.1~mJy\,beam$^{-1}$ (3$\sigma$) and increase by a factor of $-$1, 1, 2, 4, 8 and 16. }
\label{fig2}
\end{figure}

The total intensity maps of PDS~456 observed with the EVN plus the e-MERLIN on 2020 January 22 are shown in Fig.~\ref{fig2}. The beam synthesised with natural weighting has an FWHM of 23~$\times$~5.2~mas$^{2}$ in PA~=~$+73\degr$. To avoid resolving out diffuse radio features, we artificially increased the beam area and used a circular beam with a FWHM~=~23~mas in the map. The nucleus of PDS~456 consists of two relatively compact knots, denoted as the components C and NW, and a diffuse emission region extending mainly toward West and South. According to the position of its centroid, we labelled the complex structure as the component W. The unshown image from the first epoch observation has a significantly lower quality owing to the limited telescopes and sensitivity, yet it independently confirms the existence of the components C, NW and W. The relatively bright spot at $\sim$12$\sigma$ in the South, Fig.~\ref{fig2} was only marginally ($\sim$3$\sigma$) detected in the early epoch.

The circular Gaussian model fitting results in \textsc{difmap} (version 2.5e) are given in Table~\ref{tab2}. All the errors are formal uncertainties measured at the reduced $\chi^2 = 1$. Empirically, the flux density measurements have a systematic error of around ten per cent. Compared to the flux density (16~$\pm$~2~mJy at 1.66~GHz) predicted by its radio spectrum \citep{Yang2019}, the VLBI map only restores $\sim$70 per cent. With respect to J1724$-$1443, the differential astrometry gives a J2000 position of RA~=~$17^{\rm h}28^{\rm m}19\fs78958$ and Dec.~=~$-14\degr15\arcmin55\farcs8520$ for the peak component C. This is in agreement with the average position of components C1 and C2 at 5~GHz reported by \citep{Yang2019} if we consider significant systematic error due to the extended source structure and the different ($u$, $v$) coverage, $\sim$0.1$\theta_{\rm size}$ ($\sim$1~mas). With respect to the component C, we reported the relative offsets of components NW and W in Table~\ref{tab2}.           

In the last two columns of Table~\ref{tab2}, we also report radio luminosity $L_{\rm R} = \nu L_\nu$ ($\nu$, observing frequency) and brightness temperature $T_{\rm b}$, estimated \citep[e.g.,][]{Condon1982} as
\begin{equation}
T_\mathrm{b} = 1.22\times10^{9}\frac{S_\mathrm{int}}{\nu_\mathrm{obs}^2\theta_\mathrm{size}^2}(1+z),
\label{eq1}
\end{equation}
where $S_\mathrm{int}$ is the integrated flux density in mJy, $\nu_\mathrm{obs}$ is the observation frequency in GHz, $\theta_\mathrm{size}$ is the FWHM of the circular Gaussian model in mas, and $z$ is the redshift. The peak component C has a brightness temperature of $\sim$1~$\times$~10$^7$~K. The very extended component W has a brightness temperature of $\sim$3~$\times$~10$^5$~K.

\begin{table*}
\caption{Summary of characteristic parameters of the radio components detected in PDS~456. Columns give (1) component name, (2) total flux density, (3-4) relative offsets in Right Ascension and Declination with respect to the peak component C, (5) best-fit size $\theta_{\rm size}$ of the Circular Gaussian model, (6) brightness temperature $T_{\rm b}$ and (7) radio luminosity $L_{\rm R}$.  }
\label{tab2}
\begin{tabular}{ccccccc}
\hline\hline
Name & $S_{\rm int}$   & $\delta_{\rm ra}$  &  $\delta_{\rm dec}$  & $\theta_{\rm size}$ & $\log T_{\rm b}$    & $\log L_{\rm R}$   \\   
     & (mJy)           & (mas)              &  (mas)               & (mas)               & (K)                 & (erg s$^{-1}$)      \\
\hline
C    & 2.31~$\pm$~0.03 &   $+$0.0~$\pm$~0.1 &    $+$0.0~$\pm$~0.1  &   11.1~$\pm$~0.4    &  6.99~$\pm$~0.05    & 39.60~$\pm$~0.04    \\
NW   & 1.15~$\pm$~0.03 &  $-$62.7~$\pm$~0.2 &   $+$48.8~$\pm$~0.2  &   14.7~$\pm$~0.7    &  6.45~$\pm$~0.06    & 39.30~$\pm$~0.04   \\  
W    & 8.18~$\pm$~0.14 &  $-46.3$~$\pm$~0.9 &   $-$11.0~$\pm$~0.8  &  118.6~$\pm$~1.6    &  5.48~$\pm$~0.05    & 40.15~$\pm$~0.04   \\
\hline
\end{tabular}
\end{table*}

\section{Discussion}
\label{sec4}

\subsection{Star formation rate based on IR photometries}
\label{sec4-1}
To estimate the star formation rate (SFR) of PDS~456, we constructed the IR SED based on the multi-band photometries in the literature. The photometries of the Wide-field Infrared Survey Explorer ({\it WISE}), the Infrared Astronomical Satellite ({\it IRAS}), the Infrared Space Observatory ({\it ISO}) and the James Clerk Maxwell Telescope (JCMT) Submillimetre Common-User Bolometer Array 2 (SCUBA-2) have been retrieved from NASA/IPAC Extragalactic Database. The {\it Herschel} photometries have been obtained by using the Herschel Interactive Processing Environment (\textsc{hipe} v15.0.1). Compared to the previous SFR estimate based on SED fitting by \citet{Yun2004}, we add two {\it WISE} bands at 12 and 22~$\umu$m, two {\it Herschel} PACS (Photodetector Array Camera and Spectrometer) bands at 70 and 160 $\umu m$, and three SPIRE (Spectral and Photometric Imaging REceiver) bands at 250, 350 and 500~$\umu$m. The inclusion of the {\it Herschel} PACS and SPIRE photometry is crucial for the SFR estimate as they map the cold dust emission peaking at 100--200~$\umu$m, which is directly involved in the star formation activity. We use the Bayesian SED modelling and interpreting code \textsc{BAYESED} \citep{Han2012, Han2014, Han2019} to decompose the IR SED by using an AGN torus model and a simple modified blackbody model to represent the contribution from cold dust emission heated by a young stellar population \citep[for more details, see][]{Fan2016, Fan2017, Fan2019}. 

The derived IR luminosity of cold dust is $\sim~6.9~\times~10^{11}$~L$_{\sun}$, which is lower than that of \citet{Yun2004} by a factor of about two. The difference is mainly due to the following two factors. First, our SED included the more data points at far IR wavelengths than that of \citet{Yun2004}. Second, we excluded the contribution of the AGN torus emission in the IR luminosity estimate. We convert the derived IR luminosity of cold dust into a SFR of about 69~M$_{\sun}$~yr$^{-1}$, assuming a Chabrier initial mass function \citep{Chabrier2003}. We note that this SFR estimate should be taken as an upper limit due to source blending in the IR bands. The SFR is also in agreement with the results reported by \citet{Bischetti2019}, suggesting significant but no extreme star forming activity in the host galaxy. 

\subsection{Complex nuclear radio activity}
\label{sec4-2}
The sub-kpc scale radio nucleus of PDS~456 shows a complex radio morphology. This is not unusual in ultraviolet to sub-mm luminous but radio weak/quiet sources because their radio emission originate from multiple physical mechanisms including low-radio-power jets, starbursts and shocks \citep[e.g.][]{Panessa2019}. Owning to the absence of bright radio cores and the missing of sensitive short baselines, VLBI observations of the radio counterparts of these galaxies often show non-detections or significantly over-resolved radio structures. These cases involve extreme UV-selected starburst galaxies \citep{Alexandroff2012}, ULIRG IRAS~23365$+$3604 \citep{Romero2012}, hot dust-obscured galaxies \citep{Frey2016} and sub-millimetre-selected galaxies \citep{Chen2020}. 

\subsubsection{Location of the radio core}
\label{sec4-2-1}
The peak component C is the most promising component hosting the radio core of PDS~456. As the only central feature in the radio emission distribution, it is quite close to the optical centroid (J2000, R.A.~$=17^{\rm h}28^{\rm m}19\fs789380$, Dec.~$=-14\degr15\arcmin55\farcs85543$, $\sigma_{\rm p} = 0.04$~mas) reported by the \textit{Gaia} astrometry \citep[DR2,][]{Gaia2018} using a point-source model. The small offset (radius $\sim$4.5~mas) is very likely due to a double-peaked optical brightness distribution \citep[peak separation $\sim$0.22~kpc,][]{Letawe2010}. The component C might have a relatively hard spectrum. Assuming no significant flux density variability between the previous EVN observations at 5~GHz \citep{Yang2019} and the new observations at 1.6~GHz, it has a spectral index of $\alpha = -0.5~\pm~0.1$. 

\subsubsection{Evidence of a low-radio-power jet}
\label{sec4-2-2}
There exists some faint emission smoothly connecting the components C and NW. The component NW could represent the head of a jet along PA~$\sim$~$-$52$\fdg$1. With respect to the component C, the apparent jet opening angle for the component NW is $\sim$10$\fdg$7. This is about a factor of two smaller than the median value of 21$\fdg$5 found in significantly beamed AGN jets \citep{Pushkarev2017}.  Thus, the jet is unlikely to be very close to the line of sight. Assuming that the jet is close to the kinematic axis of the molecular disk \citep{Bischetti2019}, the jet viewing angle is $\theta_{\rm view} = 25\degr~\pm~10\degr$. This indicates a de-projected length of 570~$\pm$~20~pc for the component NW. 
Since there is no highly relativistic jet observed among extreme accretion systems \citep{Yang2020a}, the jet in PDS~456 might not be significantly Doppler boosted. Based on its low radio luminosity, it can be identified as a low radio power jet \citep[e.g.][]{Kunert2010, An2012}.  

The component NW has also been marginally detected as the component X2 in the previous EVN observations at 5~GHz \citep{Yang2019}. The two observations give a spectral index of $\alpha = -0.8~\pm~0.1$. Owing to the near face-on disk geometry \citep{Bischetti2019} and the southern diffuse emission, only the jet on the approaching side is clearly identified. In the VLBI image at 1.6~GHz, we failed to confirm the existence of the component X1 (relative offsets, $\delta_{\rm ra} \approx 37.4$~mas, $\delta_{\rm dec} \approx -99.4$~mas), which was tentatively detected based on the shortest baseline of Ef--Wb on the receding side in the early observations at 5~GHz. Thus, the component X1 is probably an instrumental noise peak.

\subsubsection{Diffuse component W: a relic jet or a composite of starbursts and shocks}
\label{sec4-2-3}
PDS~456 has an optically thin integrated spectrum up to $>$15~GHz \citep{Yang2019}. As the component W contributes about half of the total radio emission of PDS~456 and the components C has a relatively hard spectrum, the component W may have a similar optically thin spectrum originating from non-thermal radio activity. 

It is hard to naturally identify the component W as a jet component extending further from the component NW, as W has a diffuse radio morphology and a rather different extension direction from the existing jet, and there is no hint of a bending point. With respect to the component C, the component W shows a very wide-angle ($>$180$\degr$) extension with a size of 118~$\pm$~2~mas (366~$\pm$~6 pc) and a centroid in PA~$\sim~-103\degr$. 

The component W may be an extended relic jet component close to the line of sight. Due to the peculiar jet geometry, the projected structure does not resemble an elongated jet. Compared to the inferred jet component NW, the identification necessitates a significant change in the jet direction. Currently, this information is unavailable.  

Because of its high luminosity and extended morphology, it can not be simply explained as a single supernova or a supernova remnant, which have typical peak monochromatic luminosities only up to $L_{\rm R}$ $\sim$10$^{38.7}$~erg\,s$^{-1}$ \citep{Weiler2002} and  might only explain the brightened spot in the South. It is also hard to explain the component W as a group of supernovae and supernova remnants, i.e. starbursts. The nearby extreme starburst galaxy Arp~220 has a star-formation rate (SFR) of $\sim$220~M$_{\sun}$~yr$^{-1}$ and a radio luminosity of 4~$\times$~10$^{39}$~erg\,s$^{-1}$ \citep{Varenius2016, Varenius2019}. Compared to Arp~220, PDS~456 has at least three times lower SFR (cf. Sect.~\ref{sec4-1}). Assuming that fifty per cent of radio emission is from the starburst activity, the SFR derived based on the radio luminosity \citep[Eq. 6,][]{Bell2003} would be 690~$\pm$~70~M$_{\sun}$~yr$^{-1}$. This is inconsistent with the SED-based SFR and thus requires an additional source of energy injection.      

Powerful high-speed AGN winds and outflows can naturally produce the additional synchrotron radio emission via shocks on scales $\ga$100~pc in the nuclear high-density star-forming region \citep[e.g.][]{Zakamska2014, Nims2015}. Mildly relativistic and wide-opening-angle winds have been identified in X-ray spectroscopic observations \citep[e.g.][]{Nardini2015, Matzeu2017, Reeves2020}, with reports of multiple velocity components, up to 0.46~$c$ \citep{Reeves2018}. A possible ultraviolet outflow at 0.3~$c$ has also been reported by \cite{Hamann2018}. Thus, mildly relativistic winds can extend farther out to pc scale \citep{Reeves2020}, as also inferred from Section \ref{sec4-3} and thus might have a large impact on the whole nuclear region. Two CO(3-2) molecular outflow components were identified by \citet{Bischetti2019} in the compact nuclear region. One is a blue-shifted, high-velocity ($v \sim -800$~km\,s$^{-1}$ outflow component. The other is a low-velocity ($\la$500~km\,s$^{-1}$), high-velocity dispersion (peak: $360$~km\,s$^{-1}$) outflow component, which is $\sim$50~mas west from the quasar and spatially coincident with the centroid of the radio component W. The coincidence agrees with the shock model.

The latter scenario involving shocks due to wind interaction is now probed further to estimate the SFR. This involves driving of galactic winds and consequent feedback with the host galaxy through star formation. The relevant forces in action on cold gas clouds (temperature $\leq 10^4$ K) at pc--kpc scales include ram pressure from hot outflowing gas (in which the cold clouds are embedded) and radiation pressure from the galactic disk which are in opposition to the gravitational force due to the self-gravitating region of the disk \cite[e.g.][]{Sharma2012}; the corresponding SFR is approximated in terms of the wind velocity $v_w$ and the circular velocity of the galactic disk $v_c$ 
\begin{equation}
{\rm SFR} = (0.89~M_{\sun}~{\rm yr}^{-1})~\left(\frac{v_w}{v_c}\right)^{5/2} \left(\frac{v_c}{120~{\rm km~s}^{-1}}\right)^{25/8}. 
\end{equation}
For the above estimated $v_w$ of 360--800~km s$^{-1}$ and a fiducial value of $v_c = 120$~km\,s$^{-1}$, the SFR is 13.9--102.1~M$_{\sun}$~yr$^{-1}$. As this range is consistent with the estimated 69~M$_{\sun}$~yr$^{-1}$ from the IR observations, the latter is preferred.

If the component W is associated with the nuclear molecular outflows, and formed by the AGN winds sweeping up the high-density star-forming region, the energy conversion efficiency is $\frac{L_{\rm R}}{L_{\rm bol}} \sim $10$^{-7}$. This is two orders of magnitude lower than predicted by the model of spherical wind shocks \citep{Nims2015}. However, this might be acceptable for PDS~456 as the nuclear winds can preferentially propagate along the low-density polar directions. Our estimate is also consistent with the study of \cite{Zakamska2014} which finds a strong association between powerful outflows and the radio luminosity in radio quiet quasars, with a statistical median conversion efficiency of $\frac{L_{\rm R}}{L_{\rm bol}} \sim 10^{-6}$.

\subsection{Wind velocity and the Eddington ratio}
\label{sec4-3}

If the powerful wide-angle X-ray winds \citep{Nardini2015} are radiatively driven \citep{Matzeu2017}, PDS~456 might have a super-Eddington accretion rate. The Eddington ratio $\Gamma_e$ is the ratio of the accretion disk luminosity ($L_{\rm disk}$) to the Eddington luminosity ($L_{\rm Edd}$). For an optically thick geometrically thick disk \citep{Shakura1973} in the vicinity of the SMBH,
\begin{equation}
\label{gammae}
\Gamma_e = \frac{L_{\rm disk}}{L_{\rm Edd}} \approx \frac{1}{L_{\rm Edd}}\frac{G M_{\rm bh} \dot{M}}{2 r_{\rm bh}}	= \frac{\dot{m}}{12 \epsilon},
\end{equation}
where $M_{\rm bh}$ is the SMBH mass, $\dot{M}$ is the accretion rate and $\dot{m}$ is the accretion rate scaled in terms of the Eddington rate $\dot{M}_{\rm Edd}$, $L_{\rm Edd} = \epsilon \dot{M}_{\rm Edd} c^2$ where $\epsilon$ is the efficiency factor, $r_{\rm bh} = 6~r_{\rm G} = 6~G M_{\rm bh}/c^2$ ($r_{\rm G}$ is the gravitational radius) is the assumed distance from which the radiation is assumed to peak and corresponds to the innermost stable circular orbit for a Schwarzschild black hole (non-rotating). The evolution of velocity of a free particle driven by radiation pressure from the inner disk is given by \cite[e.g.][]{Abramowicz1990, Mohan2015} 
\begin{equation}
\label{betax}
\frac{d\beta}{dx} = \frac{(1-\beta^2)}{2 x^2 \beta \xi} \left(\frac{\Gamma_e \xi^{1/2}_s}{(1-\beta^2)^{1/2} \xi^{1/2}} \left(1+\beta^2-\frac{8}{3}\beta\right)-1\right),
\end{equation}
where $x = r/r_{\rm G}$, $\xi = (1-1/x)$, $\xi_s = \xi(x = 6)$ and we have assumed a radial particle trajectory which is shown to be the case for the large distance regime \citep{Mohan2015}. The formulation corresponds to a wide-angle wind outflow launched from the innermost region subject to a radiative deceleration (an effective drag force), aided by the gravitational potential of the black hole. This can be used to estimate the Eddington ratio corresponding to an observed asymptotic wind velocity at the sub-pc -- pc-scale. 

The equation~(\ref{betax}) is solved assuming $\epsilon = 0.1$ and $\Gamma_e = 0.1 - 10.0$ (sub -- super Eddington luminosity) which corresponds to $\dot{m} = 0.12 - 12.0$ using equation~(\ref{gammae}). For a Keplerial angular velocity $\Omega_{\rm bh} = G M_{\rm bh}/r^3_{\rm bh}$ associated with the launch radius $r_{\rm bh}$, the rotational velocity $v_\phi = r_{\rm bh} \Omega_{\rm bh} = c/6^{1/2} \approx$ 0.41~$c$. This could represent the minimum velocity of disk material that is advected into the winds and possibly into the collimated jet; for the boundary condition associated with equation~(\ref{betax}), we assume an initial launch velocity $\beta (r_{\rm bh}) = 0.41$. A launch velocity exceeding 0.41~$c$ could indicate a poloidal component away from the disk plane and hint at mechanisms driving the material into trajectories away from the canonical Keplerian disk orbits, including a truncated disk scenario with a magnetically arrested disk accretion in the innermost region \cite[e.g.][]{Tchekhovskoy2011} and energy extraction from the spin of the black hole \citep{Blandford1977, Blandford1982}. 

\begin{figure}
\centerline{\includegraphics[width=\columnwidth]{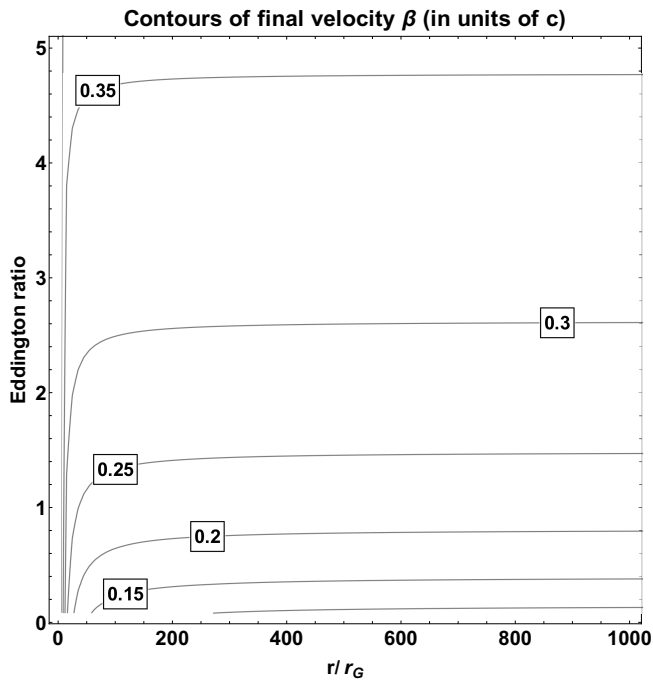}}
\caption{Contours of velocity $\beta$ from equation~(\ref{betax}) as a function of the Eddington ratio and the distance from the SMBH $r/r_{\rm G}$ (up to 1000, about 0.1~pc). The $\beta$ values tend to a constant before about 0.1~pc for an initial launch velocity of 0.41~$c$. The radiatively driven and mildly relativistic winds with $\beta$~=~0.1--0.3~$c$ indicate the Eddington ratio $\Gamma$~=~0.2--2.6.}
\label{figbetax}
\end{figure}

The above formulation can be generalized to estimate the velocity (or Lorentz factor) in the launching region for sources which indicate accretion rates near or exceeding the Eddington rate. The contours of the asymptotic velocity $\beta (\Gamma_e, x)$ are plotted in Fig.~\ref{figbetax}. The particle velocity tends to a constant by $\approx 1000 ~r_{\rm G} \approx 0.1$~pc (in the AGN rest frame). With an increase in $\Gamma_e$, the velocity saturates to a larger value for the assumed $\beta (r_{\rm bh})$. Wind velocities of $\leq$
0.3~c,  as suggested by the radio and X-ray observations can originate from the inner disk. These can be achieved well in advance and stay at the saturated value up to the pc-scale and possibly beyond. The Eddington ratio corresponding to $\beta$ $\leq$ 0.3~$c$ is $\Gamma_e \leq 2.6$ 
($\dot{M} \leq$ 71.3~M$_{\sun}$~yr$^{-1}$), indicating that PDS~456 likely accretes at rates exceeding the Eddington rate. The estimated $\Gamma_e \leq 2.6$ suggests that the bolometric luminosity \citep{Reeves2000} may have been slightly underestimated (upto $3 \times 10^{47}$ erg s$^{-1}$) or the SMBH mass \citep{Nardini2015} may have been overestimated ($M_{\rm bh}$ as low as $3 \times 10^8~M_\odot$). 

The formulation can act an independent manner of constraining the Eddington ratio (and the associated accretion rate) in this and similar super-Eddington sources using the observed wind velocity as an input.

\section{Conclusions}
\label{sec5}
The luminous radio-quiet quasar PDS~456 is a nearby extreme accretion and feedback system. With the EVN plus e-MERLIN observations at 1.66~GHz, we achieved a more complete view of its complex radio morphology. Our VLBI images revealed a complex radio-emitting nucleus consisting of a collimated jet and a very extended structure. The latter has a size of about 120~mas and a radio luminosity about three times higher than the nearby extreme starburst galaxy Arp~220. The powerful nuclear radio activity could be explained as a relic jet with an unusual jet geometry, or a more natural consequence of star-forming activity plus AGN-driven wind shocks. In addition, fainter nuclear radio activity and a sub-mJy flat-spectrum jet base may exist. The observed relativistic X-ray winds of $\sim$0.3~$c$ in the inner sub-pc-scale region suggests that PDS~456 likely accretes exceeding the Eddington rate. Future multi-frequency deep VLBI observations of PDS~456 can potentially provide more details on the complex nuclear activity.

\section*{Acknowledgements}
This work was partly supported by the Square Kilometer Array pre-research funding from the Ministry of Science and Technology of China (2018YFA0404603) and the Chinese Academy of Sciences (CAS, No. 114231KYSB20170003).
EN acknowledges financial support from the agreement ASI-INAF n.2017-14-H.0.
LF acknowledges the support from National Natural Science Foundation of China (NSFC, Grant No. 11822303, 11773020 and 11421303) and Shandong Provincial Natural Science Foundation, China (JQ201801).
The European VLBI Network (EVN) is a joint facility of independent European, African, Asian, and North American radio astronomy institutes. Scientific results from data presented in this publication are derived from the following EVN project code(s): EY027. 
e-VLBI research infrastructure in Europe is supported by the European Union’s Seventh Framework Programme (FP7/2007-2013) under grant agreement number RI-261525 NEXPReS.
e-MERLIN is a National Facility operated by the University of Manchester at Jodrell Bank Observatory on behalf of STFC.
The research leading to these results has received funding from the European Commission Horizon 2020 Research and Innovation Programme under grant agreement No. 730562 (RadioNet).
This work has made use of data from the European Space Agency (ESA) mission {\it Gaia} (\url{https://www.cosmos.esa.int/gaia}), processed by the {\it Gaia} Data Processing and Analysis Consortium (DPAC, \url{https://www.cosmos.esa.int/web/gaia/dpac/consortium}). Funding for the DPAC has been provided by national institutions, in particular the institutions participating in the {\it Gaia} Multilateral Agreement. 
This research has made use of the NASA/IPAC Extragalactic Database (NED), which is operated by the Jet Propulsion Laboratory, California Institute of Technology, under contract with the National Aeronautics and Space Administration.
This research has made use of NASA’s Astrophysics Data System Bibliographic Services.

\section*{Data Availability}
The correlation data of the EVN experiment EY027 are available to the public in the EVN data archive. The calibrated visibility data and the final image files underlying this article will be shared on reasonable request to the corresponding author.





\begin{thebibliography}{99}
\bibitem[\protect\citeauthoryear{Abramowicz, Ellis \& Lanza}{1990}]{Abramowicz1990}
Abramowicz M.~A., Ellis G.~F.~R., Lanza A., 1990, \apj, 361, 470
\bibitem[\protect\citeauthoryear{Alexandroff et al.}{2012}]{Alexandroff2012}
Alexandroff R., et al., 2012, \mnras, 423, 1325 
\bibitem[\protect\citeauthoryear{An \& Baan}{2012}]{An2012}
An T., Baan W.~A., 2012, \apj, 760, 77
\bibitem[\protect\citeauthoryear{An et al.}{2013}]{An2013}
An T., Baan W.~A., Wang J.-Y., Wang Y., Hong X.-Y., 2013, \mnras, 434, 3487
\bibitem[\protect\citeauthoryear{Bell}{2003}]{Bell2003}
Bell E.~F., 2003, \apj, 586, 794
\bibitem[\protect\citeauthoryear{Bischetti et al.}{2019}]{Bischetti2019}
Bischetti M., et al., 2019, \aap, 628, A118
\bibitem[\protect\citeauthoryear{Blandford, Meier \& Readhead}{2019}]{Blandford2019} 
Blandford R., Meier D., Readhead A., 2019, \araa, 57, 467
\bibitem[\protect\citeauthoryear{Blandford \& Payne}{1982}]{Blandford1982}
Blandford R.~D., Payne D.~G., 1982, MNRAS, 199, 883
\bibitem[\protect\citeauthoryear{Blandford \& Znajek}{1977}]{Blandford1977}
Blandford R.~D., Znajek R.~L., 1977, MNRAS, 179, 433
\bibitem[\protect\citeauthoryear{Boissay-Malaquin et al.}{2019}]{Boissay-Malaquin2019}
Boissay-Malaquin R., Danehkar~A., Marshall H.~L., Nowak M~A., 2019, \apj, 873,29
\bibitem[\protect\citeauthoryear{B\"ottcher, Harris \& Krawczynski }{2012}]{Bottcher2012}
B\"ottcher M., Harris D.~E., Krawczynski H., 2012, Relativistic
Jets from Active Galactic Nuclei. Wiley, Weinheim
\bibitem[\protect\citeauthoryear{Briggs}{1995}]{Briggs1995}
Briggs D.~S., 1995, \baas, 27, 1444
\bibitem[\protect\citeauthoryear{Chabrier}{2003}]{Chabrier2003}
Chabrier G., 2003, \pasp, 115, 763
\bibitem[\protect\citeauthoryear{Chen et al.}{2020}]{Chen2020}
Chen H., et al., 2020, \aap, 638, A113
\bibitem[\protect\citeauthoryear{Condon et al.}{1982}]{Condon1982}
Condon J.~J., Condon M.~A., Gisler G., Puschell J.~J., 1982, \apj, 252, 102
\bibitem[\protect\citeauthoryear{Fan et al.}{2016}]{Fan2016}
Fan L., Han Y., Nikutta R., Drouart G., Knudsen K.~K., 2016, \apj, 823, 107
\bibitem[\protect\citeauthoryear{Fan et al.}{2017}]{Fan2017}
Fan L., Jones S.~F., Han Y., Knudsen K.~K., 2017, \pasp, 129, 124101 
\bibitem[\protect\citeauthoryear{Fan et al.}{2019}]{Fan2019}
Fan L., Knudsen K.~K., Han Y., Tan Q.-H., 2019, \apj, 887, 74
\bibitem[\protect\citeauthoryear{Fiore et al.}{2017}]{Fiore2017}
Fiore F., et al., 2017, \aap, 601, A143
\bibitem[\protect\citeauthoryear{Frey et al.}{2016}]{Frey2016}
Frey S., Paragi Z., Gab\'anyi K.~\'E., An T., 2016, \mnras, 455, 2058
\bibitem[\protect\citeauthoryear{\textit{Gaia} Collaboration}{2018}]{Gaia2018}
\textit{Gaia} Collaboration, 2018, \aap, 616, A1
\bibitem[\protect\citeauthoryear{Giroletti et al.}{2017}]{Giroletti2017}
Giroletti M., Panessa F., Longinotti A.~L., Krongold Y., Guainazzi M., Costantini E., Santos-Lleo M., 2017, \aap, 600, A87
\bibitem[\protect\citeauthoryear{Greisen}{2003}]{Greisen2003} 
Greisen E.~W., 2003, in Heck A., ed., Astrophysics and Space Science Library, Vol. 285, Information Handling in Astronomy: Historical Vistas. Kluwer, Dordrecht, p. 109
\bibitem[\protect\citeauthoryear{Hamann et al.}{2018}]{Hamann2018}
Hamann F., Chartas G., Reeves J., Nardini E., 2018, \mnras, 476, 943
\bibitem[\protect\citeauthoryear{Han \& Han}{2012}]{Han2012}
Han Y., Han Z., 2012, \apj, 749, 123
\bibitem[\protect\citeauthoryear{Han \& Han}{2014}]{Han2014}
Han Y., Han Z., 2014, \apjs, 215, 2
\bibitem[\protect\citeauthoryear{Han \& Han}{2019}]{Han2019}
Han Y., Han Z., 2019, \apjs, 240, 3
\bibitem[\protect\citeauthoryear{Keimpema et al.}{2015}]{Keimpema2015}
Keimpema A., et al., 2015, Exp. Astron., 39, 259
\bibitem[\protect\citeauthoryear{Kettenis et al.}{2006}]{Kettenis2006}
Kettenis M., van Langevelde H.~J., Reynolds C., Cotton B., 2006, Astron. Data Anal.
Software Syst. XV, 351, 497
\bibitem[\protect\citeauthoryear{Kormendy \& Ho}{2013}]{Kormendy2013}
Kormendy J., Ho L.~C., 2013, \araa, 51, 511
\bibitem[\protect\citeauthoryear{Kunert-Bajraszewska et al.}{2010}]{Kunert2010}
Kunert-Bajraszewska M., Gawro\'nski M.~P., Labiano A., Siemiginowska A.,  2010, \mnras, 408, 2261
\bibitem[\protect\citeauthoryear{Lamastra et al.}{2017}]{Lamastra2017}
Lamastra A., Menci N., Fiore F., Antonelli L.~A., Colafrancesco S., Guetta D., Stamerra A., 2017, \aap, 607, A18
\bibitem[\protect\citeauthoryear{Letawe, Letawe \& Magain}{2010}]{Letawe2010}
Letawe Y., Letawe G., Magain, P., 2010, \mnras, 403, 2088
\bibitem[\protect\citeauthoryear{Matzeu et al.}{2017}]{Matzeu2017}
Matzeu G.~A., Reeves J.~N., Braito V., Nardini E., McLaughlin D.~E., Lobban A.~P., Tombesi F., Costa M.~T., 2017, \mnras, 472, L15
%
\bibitem[\protect\citeauthoryear{Mohan \& Mangalam}{2015}]{Mohan2015}
Mohan P., Mangalam A., 2015, \apj, 805, 91
\bibitem[\protect\citeauthoryear{Nardini et al.}{2015}]{Nardini2015}
Nardini E., et al., 2015, \sci, 347, 860
\bibitem[\protect\citeauthoryear{Nims, Quataert \& Faucher-Gigu\`ere}{2015}]{Nims2015}
Nims J., Quataert E., Faucher-Gigu\`ere C.-A., 2015, \mnras, 447, 3612
\bibitem[\protect\citeauthoryear{O'Brien et al.}{2005}]{OBrien2005}
O'Brien P.~T., Reeves J.~N., Simpson C., Ward M.~J., 2005, \mnras, 360, L25
\bibitem[\protect\citeauthoryear{Panessa et al.}{2019}]{Panessa2019}
Panessa F., Baldi R.~D., Laor A., Padovani P., Behar E., McHardy I. 2019, Nat. Astron., 3, 387
\bibitem[\protect\citeauthoryear{Petrov et al.}{2006}]{Petrov2006}
Petrov~L., Kovalev Y.~Y., Fomalont E.~B., Gordon D., \aj, 2006, 131, 1872
\bibitem[\protect\citeauthoryear{Pushkarev et al.}{2017}]{Pushkarev2017}
Pushkarev A.~B., Kovalev K.~K., Lister M.~L., Savolainen T., 2017, \mnras, 468, 4992
\bibitem[\protect\citeauthoryear{Reeves et al.}{2000}]{Reeves2000} 
Reeves J.~N., O'Brien P.~T., Vaughan S., Law-Green D., Ward M., Simpson C., Pounds K.~A., Edelson R., 2000, \mnras, 312, L17
\bibitem[\protect\citeauthoryear{Reeves et al.}{2018}]{Reeves2018} 
Reeves J.~N., Braito V., Nardini E., Lobban A.~P., Matzeu G.~A., Costa M.~T.,  2018, \apj, 854, L8
\bibitem[\protect\citeauthoryear{Reeves et al.}{2020}]{Reeves2020} 
Reeves J.~N., Braito V., Chartas G., Hamann F., Laha S., Nardini E., 2020, \apj, 895, 37
\bibitem[\protect\citeauthoryear{Romero-Ca\~nizales, P\'erez-Torres \& Alberdi}{2012}]{Romero2012}
Romero-Ca\~nizales C., P\'erez-Torres M.~A., Alberdi A., 2012, 422, 510
\bibitem[\protect\citeauthoryear{Sharma \& Nath}{2012}]{Sharma2012}
Sharma M., Nath B.~B., 2012, \apj, 750, 55
\bibitem[\protect\citeauthoryear{Shakura \& Sunyaev}{1973}]{Shakura1973}
Shakura N.~I., Sunyaev R.~A., 1973, \aap, 500, 33
\bibitem[\protect\citeauthoryear{Shepherd, Pearson \& Taylor}{1994}]{Shepherd1994} 
Shepherd M.~C., Pearson T.~J., Taylor G.~B., 1994, \baas, 26, 987
\bibitem[\protect\citeauthoryear{Simpson et al.}{1999}]{Simpson1999}
Simpson C., Ward M., O'Brien P., Reeves J., 1999, \mnras, 303, L23
\bibitem[\protect\citeauthoryear{Tchekhovskoy, Narayan \& McKinney}{2011}]{Tchekhovskoy2011}
Tchekhovskoy A., Narayan R., McKinney J.~C., 2011, \mnras, 418, L79
\bibitem[\protect\citeauthoryear{Tombesi et al.}{2015}]{Tombesi2015}
Tombesi F., Mel\'endez M., Veilleux S., Reeves J.~N., Gonz\'alez-Alfonso E., Reynolds C.~S., 2015, \nat, 519, 436
\bibitem[\protect\citeauthoryear{Tombesi}{2016}]{Tombesi2016}
Tombesi F., 2016, Astron. Nachr., 337, 410
\bibitem[\protect\citeauthoryear{Torres et al.}{1997}]{Torres1997}
Torres C.~A.~O., Quast G.~R., Coziol R., Jablonski F., de la Reza R., L\'epine J.~R.~D., Greg\'orio-Hetem J., 1997, \apjl, 488, L19
\bibitem[\protect\citeauthoryear{Varenius et al.}{2016}]{Varenius2016}
Varenius E., et al., 2016, \aap, 593, A86
\bibitem[\protect\citeauthoryear{Varenius et al.}{2019}]{Varenius2019}
Varenius E., et al., 2019, \aap, 623, A173
\bibitem[\protect\citeauthoryear{Weiler et al.}{2002}]{Weiler2002}
Weiler K.~W., Panagia N., Montes M.~J., Sramek R.~A., 2002, \araa, 40, 387    
\bibitem[\protect\citeauthoryear{Yang et al.}{2019}]{Yang2019}
Yang J., An T., Zheng F., Baan W.~A.; Paragi Z.; Mohan P.; Zhang Z.; Liu X., 2019, \mnras, 482, 1701
\bibitem[\protect\citeauthoryear{Yang et al.}{2020a}]{Yang2020a}
Yang J., Paragi Z., An T., Baan W.~A., Mohan P., Liu X., 2020, \mnras, 494, 1744
\bibitem[\protect\citeauthoryear{Yang et al.}{2020b}]{Yang2020b}
Yang J., Gurvits L.~I., Paragi Z., Frey S., Conway J.~E., Liu X., Cui L., 2020, \mnras, 495, L71
\bibitem[\protect\citeauthoryear{Yun et al.}{2004}]{Yun2004}
Yun M.~S., Reddy N.~A., Scoville N.~Z., Frayer D.~T., Robson E.~I., Tilanus R.~P.~J., 2004, ApJ, 601, 723
\bibitem[\protect\citeauthoryear{Zakamska \& Greene}{2014}]{Zakamska2014}
Zakamska N.~L., Greene J.~E., 2014, \mnras, 442, 784  



\end{thebibliography}

%



\appendix




\bsp	
\label{lastpage}
\end{document}